\documentstyle{article}
\begin{document}
\openup6pt
\title {New static spheroidal solution in Jordan-Brands-Dicke theory.}
\author{S.M.KOZYREV \thanks{Email address: Sergey@tnpko.ru}  \\
Scientific center gravity wave studies ''Dulkyn'', \\
PB 595, Kazan, 420111, Russia ,Kazan, Russian Federation \\}
\date{}
\maketitle
\begin{abstract}
The static spheroidal solutions of Jordan-Brands-Dicke theory
(JBD) are studied. We consider the effect of the anisotropic
stresses of scalar field on the shape of JBD self-graviting
objects. It is shown that scalar fields can have significant
effect on the structure and properties of self-graviting objects.
In contrast with general relativity in JBD theory there are
nonflat static spheroidal solutions.
\end{abstract}
\section{Introduction}
A wide spread assumption in the study of stellar structure is that
the shape of star can be modeled as a spherical symmetry object.
This approach has been used extensively in the study of star, star
system and galaxies \cite{Finch}. However, in many systems,
deviation from spherical symmetry may play an important role in
determining of them properties. Physical situation where
unspherical shape may be relevant are very diverse. Scalar
self-graviting objects resulting from the non minimal coupling
scalar fields to gravity are a system where anisotropic pressure
occurs naturally \cite{Lee}. A model for the Universe where the
dark matter and energy are the scalar nature can be realistic and
could explain most of the observed structures \cite{Will}.
 The self-interaction of the scalar field could explain the behavior
 of galaxy rotation curves all along the background.

 Anisotropy appears as an extra assumption on the
behavior of scalar fields and on the shape of equilibrium
configuration. Since we still do not have a formulation of the
possible anisotropic stresses is emerging in these or other
contexts, we take the approach of finding several exact solutions
representing physical situations, modelled by ellipsoid of
revolution. Our goals hear is to find exact spheroidal solution,
offering an analysis of the change in the physical properties of
the stellar and galaxy models due to presence of non minimally
coupled scalar fields. In this context, particularly interesting
is the case of JBD theory \cite{Jordan}, \cite{Brans} where
pressure anisotropies come in action.

\section{Static spheroidal vacuum solutions of Jordan-Brans-Dicke theory.}

JBD theory can be thought of as a minimal extension of general
relativity designed to properly accommodate both Mach's principle
and Dirac's large number hypothesis. The progress in the
understanding of scalar-tensor theories of gravity is closely
connected with finding and investigation of exact solutions.
Shortly after JBD theory was proposed, Heckmann obtained
parametric form of the exact static vacuum solution to the JBD
equations \cite{Heckmann}. Later Brans \cite{Brans} find the
static, spherically symmetric, vacuum solution of the JBD
equations in isotropic coordinates.
    In the Jordan conformal frame, the JBD action takes the form \cite{Jordan}
    (we use geometrized units such that G = c = 1 and we follow the signature +,-,-,-).

\begin{eqnarray}
S &=&\int dx\sqrt{-g}(\phi R-\omega  g^{\mu \nu }\nabla _\mu \phi
\nabla _\nu \phi )+S_m. \label{eq3}
\end{eqnarray}

Here, $\omega $ coupling constant, \textit{R} is the Ricci scalar
curvature with respect to the space-time metric g$_{\mu \nu }$ and
S$_m$ denotes action of matter fields. (We use units in which
gravitational constant \textit{G}=1 and speed of light c=1.)

Variation of (\ref{eq3}) with respect to g$_{\mu \nu }$ and $\phi
$ gives, respectively, the field equations:
\begin{equation}
R_{\mu \nu }-\frac 12Rg_{\mu \nu }=\frac 1 {\phi \ }T_{\mu \nu
}^M+ T_{\mu \nu }^{JBD},  \label{eq4}
\end{equation}
where
\begin{eqnarray}
T_{\mu \nu }^{JBD} &=& [ \frac \omega {\phi ^2}\left( \nabla _\mu
\phi \nabla _\nu \phi -\frac 12g_{\mu \nu }\nabla _\alpha \phi
\nabla ^\alpha \phi \right) +  \nonumber \\[0.01in]
&&\ \ +\frac 1\phi \left( \nabla _\mu \nabla _\nu \phi -g_{\mu \nu
}\nabla _\alpha \nabla ^\alpha \phi \right) ],  \label{eq5}
\end{eqnarray}
and
\begin{equation}
\nabla _\alpha \nabla ^\alpha \phi =\frac{T_\lambda ^{M\ \lambda
}}{3 + 2 \omega },  \label{eq6}
\end{equation}
and $T_\lambda ^{M\ \lambda }$ is the energy momentum tensor of
ordinary matter which obeys the conservation equation $T_{\mu \nu
;\lambda }^{M\ }$ g$ ^{\nu \lambda }$= 0. One can derive the
energy density and pressure of JBD field from (\ref{eq5}) $ \rho
^{JBD} = T_{00}$, $ P ^{JBD} = T_{ii}$ $(i = 1,2,3)$.

As we have already mentioned we consider standard static
spheroidal space-time. Solutions to the equation in spheroidal
coordinates have application to a wide range of problems in
physics \cite {Flammer}. The two-dimensional elliptic coordinate
system is defined from the set of all ellipses and all hyperbolas
with a common set of two focal points. Oblate spheroidal
coordinates are derived from elliptic coordinates by rotating the
elliptical coordinate system about the perpendicular bisector of
the focal points. Similarly, one can obtain the prolate spheroidal
coordinates by rotating it about the parallel bisector. We adopt
coordinates that allow us to write spheroidal geometry in prolate
form
\begin{eqnarray}
ds^2=-B \left( \xi\right) dt^2+  A \left( \xi\right) \left( c
 \sqrt{\frac{\xi^2-\eta^2}{\xi^2-1 } } d \xi ^2+c
 \sqrt{\frac{\xi^2-\eta^2}{1 - \eta^2} } d\eta^2+c
 \sqrt{(\xi^2-1)(1-\eta^2)} d\varphi^2 \right),
\label{eqSphOb}
\end{eqnarray}
and in oblate form
\begin{eqnarray}
ds^2=-B \left( \xi\right) dt^2+  A \left( \xi\right) \left( c
 \sqrt{\frac{\xi^2-\eta^2}{\xi^2-1 } } d \xi ^2+c
 \sqrt{\frac{\xi^2-\eta^2}{1 - \eta^2} } d\eta^2+c
 \xi \eta d\varphi^2 \right),
\label{eqSphPr}
\end{eqnarray}
where A, B are function of $\xi$ and $\xi \geq  1$,  $-1 \leq \eta
\leq 1$, $0 \leq \varphi \leq 2\pi$. We denote the separation of
the two focal points by $c$.

 Then the solutions of the gravitational field
equations in the vacuum $T_{\mu \nu }^{M\ }$= 0 take in oblate
case the form
\begin{equation}
A=a_0 \left( \frac{1+ \sqrt{1-\xi^2}}{-1+
\sqrt{1-\xi^2}}\right)^{\beta \left( 1+
\frac{1}{\sqrt{-3-2\omega},
 }\right)}, \nonumber
\end{equation}
\begin{equation}
B =b_0 \left( \frac{1+ \sqrt{1-\xi^2}}{-1+
\sqrt{1-\xi^2}}\right)^{\beta \left( -1+
\frac{1}{\sqrt{-3-2\omega},
 }\right)}, \label{e4}\\
\end{equation}
\begin{equation}
\phi =\phi _0\left( \frac{1+ \sqrt{1-\xi^2}}{-1+
\sqrt{1-\xi^2}}\right)^{- \frac{\beta }{\sqrt{-3-2\omega},
 }}, \nonumber
\end{equation}
where $\beta, a_0, b_0, \phi _0$ arbitrary constants.
For the prolate case
\begin{equation}
A=a_0 \left( \frac{1+ \xi}{-1+ \xi}\right)^{\beta \left( \frac
{1-\sqrt{-3-2\omega}}{1+ \sqrt{-3-2\omega}}\right) }, \nonumber
\end{equation}
\begin{equation}
B =b_0 \left( \frac{1+ \xi}{-1+ \xi}\right)^{\beta  }, \label{e5}\\
\end{equation}
\begin{equation}
\phi =\phi _0\left( \frac{1+ \xi}{-1+ \xi}\right)^{\left( \frac
{-\beta }{1+ \sqrt{-3-2\omega}}\right) }. \nonumber
\end{equation}

These solutions can take some possible forms, depending on the
values of arbitrary constants appearing in the solution. Now
choose the value $\omega = - 2$ and redefine the arbitrary
constant. Then the metric and scalar function become
\begin{equation}
A=1, \nonumber
\end{equation}
\begin{equation}
B =b_0 \left( \frac{1+ \sqrt{1-\xi^2}}{-1+
\sqrt{1-\xi^2}}\right)^{\beta }, \label{e4}\\
\end{equation}
\begin{equation}
\phi =\phi _0\left( \frac{1+ \sqrt{1-\xi^2}}{-1+
\sqrt{1-\xi^2}}\right)^{- \frac{\beta }{2,
 }}, \nonumber
\end{equation}
and
\begin{equation}
A=1, \nonumber
\end{equation}
\begin{equation}
B =b_0 \left( \frac{1+ \xi}{-1+ \xi}\right)^{\beta  }, \label{e5}\\
\end{equation}
\begin{equation}
\phi =\phi _0\left( \frac{1+ \xi}{-1+ \xi}\right)^{- \frac {\beta
}{2} }. \nonumber
\end{equation}

To see that all these metrics is asymptotically flat it is enough
to show that the metric components behave in an appropriate way at
large $\xi$-coordinate values, e.g., $g_{\mu \nu } = \eta_{\mu \nu
}+ O(1/\xi)$ as $\xi \rightarrow \infty $. By inspection of the
coefficients, we verify that this is so.

It is certainly true that any vacuum solution of Einstein's
equations is also a solution of JBD equations (\ref{eq4}) with
$\phi$ strictly constant, and that $\phi=const$ is the solution of
the equation(\ref{eq6}) for $\omega  \rightarrow \infty $.
However, this by no means implies that all JBD solutions satisfy
Einstein's equations in the limit  $\omega \rightarrow \infty $ or
in  the limit $\phi \rightarrow const $. In fact, it is easy to
show that Einstein's field equations yield only the flat space
$A=1$, $B=1$.

We now deduce the results obtained in the oblate and prolate
spheroidal coordinates. Firstly, in both cases, the JBD scalar
field appears to play the role of dark matter component. Secondly,
the directional components of equation of state of JBD field are
anisotropic and in both cases indicate that the JBD scalar field
appears to be an "exotic" type of dark matter.

\section{Discussion}
In this article we delineated the qualitative features one would
expect from spheroidal scalar field object. It is demonstrated
that our model  can successfully predict the spheroidal
configuration in terms of a self-gravitating spacetime solution to
the JBD field equations and reproduce the not
spherically-symmetric shape in terms of the non-trivial energy
density and anisotropic pressure of the JBD scalar field which was
absent in the context of general relativity. We believe that
following this hypotheses the shape of galaxy and rotation curve
may be explained by action of scalar fields. The solution
presented here could be a first approximation at the galactic
space-time provided the presence of scalar field. Therefore, it is
necessary to study how these results modify the standard method of
interpretation rotation data. Further investigation into the
nature solutions with view to separating the real rotational
effects from the scalar fields anisotropy might be rewarding.

\end{document}